\documentclass[epsf,twocolumn,preprintnumbers]{revtex4}
\usepackage{graphics}
\usepackage{graphicx}
\usepackage{dcolumn} 
\usepackage{bm}
\usepackage{epsfig}
\begin{document}

\title{Comparison of the Electronic Structures of Two Non-cuprate\\
    Layered Transition Metal Oxide Superconductors}
\author{E. R. Ylvisaker, K.-W. Lee, and W. E. Pickett}
\affiliation{Department of Physics, University of California, Davis, 
   California, 95616}
\date{\today}

\begin{abstract}
Comparison is made of the electronic structure
of the little-studied layered transition metal oxide LiNbO$_2$ with 
that of Na$_x$CoO$_2$, which has
attracted tremendous interest since superconductivity 
was discovered in its hydrate.
Although the active transition metal $d$ states are quite different
due to different crystal fields and band filling, both systems show a strong
change of electronic structure with changes in the distance between the
transition metal ion layer and the oxygen layers.  The
niobate is unusual in having a large second-neighbor hopping amplitude, 
and a nearest neighbor hopping amplitude that is sensitive to the Nb-O
separation.  
Li$_x$NbO$_2$ also presents the attractive simplicity of a single band 
triangular lattice system with variable carrier concentration that is
superconducting.
\end{abstract}
\maketitle

\section{Motivation}
Among the various areas of research that were stimulated by the discovery
of high temperature superconductors (HTS) nearly two decades
ago is that of two-dimensional
(2D) (or nearly so) transition metal oxides (TMOs).  A second surprise appeared
in 2001 with the discovery\cite{akimitsu} of T$_c$ = 40 K in MgB$_2$,
where the physics is entirely different but the 2D character is 
crucial\cite{mazin,pickett}
for the surprisingly high value of critical temperature T$_c$.  A further
stimulus for study of superconductivity in 2D TMOs was provided in 2003
with the discovery of superconductivity\cite{takada} in hydrated
Na$_x$CoO$_2$ at 4.5 K.  These discoveries suggest a more general look at
superconducting 2D TMOs besides the cuprates, to try to identify trends
(or perhaps lack of trends). 

Being isostructural to the first HTS (La,Sr)$_2$CuO$_4$, the ruthenate
Sr$_2$RuO$_4$ has a special status in this class.  Its electronic structure
is quite distinct from that of HTS, however, and T$_c$ is only around 1 K.
There is now a very large literature on Sr$_2$RuO$_4$. It is a different
and very perplexing superconductor, but we will not pursue it in this
paper.

What we focus on here is the little-noticed layered TMO superconductor
Li$_x$NbO$_2$, with brief comparison with the cobaltate system
Na$_x$CoO$_2$.  This niobate was discovered\cite{Geselbract-Nature} 
in 1990 when the community
was absorbed with the new HTS materials, and has not yet attracted the
attention that it deserves.  While its T$_c$ = 5.5 K is quite close to
that of the hydrated cobaltates (4.5 K), it is the contrasts that we will
focus on.  These differences revolve mainly on: $4d$ versus $3d$ ion,
trigonal versus octahedral coordination by six oxygen neighbors, and
single band versus multiband character.  We expose one similarity:
$z$-displacement of the oxygen layers, which modulates the TM-oxygen
distance, has a strong influence on the electronic structure.

\section{Layered Lithium Niobate}
The compound LiNbO$_2$ itself is a band insulator with gap $\sim$2 eV.
The de-lithiated phase Li$_x$NbO$_2$ was found by the Berkeley group to
be superconducting,\cite{Geselbract} with the few reports to date
suggesting superconductivity sets in around $x\approx 0.8$ ({\it i.e.}
when 20\% of the Li is removed), beyond which T$_c$ does not 
depend much on the Li content $x$.
The structure of LiNbO$_2$ consists of a triangular lattice of 
both the Li cations and the transition
metal (niobium) ions, separated by layers of oxygen atoms, 
similar to Na$_x$CoO$_2$ except for the TM coordination. 
The trigonal prismatic
coordination of niobium atoms by oxygen ions provides a big 
distinction.  The trigonal 
crystal field results in an energetic lowering of the Nb d$_{z^2}$
states with respect to the other $4d$ states by about 4 eV, leaving the 
system with only a single band per formula
unit to consider.  This valence-conduction band is also well
separated from the O $2p$ bands below (see Fig. 1).

Removal of the lithium has the effect
of adding holes to the conduction band made up of Nb $d_{z^2}$ states.  
Superconductivity appears, as it does when holes are introduced
into NaCoO$_2$ (followed by hydration), and at a very similar
temperature (5 K), but apparently at quite different carrier
concentrations and for very different electronic structures.
Since the Li content is variable, this compound becomes a clean
representation of a single band triangular lattice system which
can be compared rather directly with Hubbard model results.  As part
of our study of this system, we obtain a tight-binding (TB) 
representation of the band to allow the subsequent study of possible
correlation effects within the Hubbard model.  We return to these
issues below.

{\it Structure.}
LiNbO$_2$ takes on a hexagonal 
structure\cite{Mosh,Geselbract,Meyer} ($a$=2.90~\AA, 
$c$=10.46~\AA) having
space group $P6_3/mmc$ (No. 194), with sites Li 
[$2a$ (0,0,0), ${\overline 3}m$], Nb 
[$2d$ $(\frac23, \frac13, \frac14)$, ${\overline 6}m2$], and O
[$4f$ $(\frac13, \frac23, z_O)$, $3m$].  The oxygen  
internal parameter $z_O$ specifies the Nb-O bond length, and due to
the stacking type there are two LiNbO$_2$ layers per cell.
The distance between Nb atoms, $a$, 
is almost identical to bond length 2.86~\AA~in elemental bcc Nb,
so direct Nb-Nb overlap should be kept in mind.  Experimental 
values\cite{Meyer,Mosh,Geselbract,Tyut} of the internal parameter 
range from 0.125-0.129.  Our optimization by energy minimization
using the abinit code gives the value $z_O$=0.125 
(lattice constants held at the 
experimental values).

\begin{figure}[bt]
\label{fig:Band}
\resizebox{8cm}{5cm}{\includegraphics{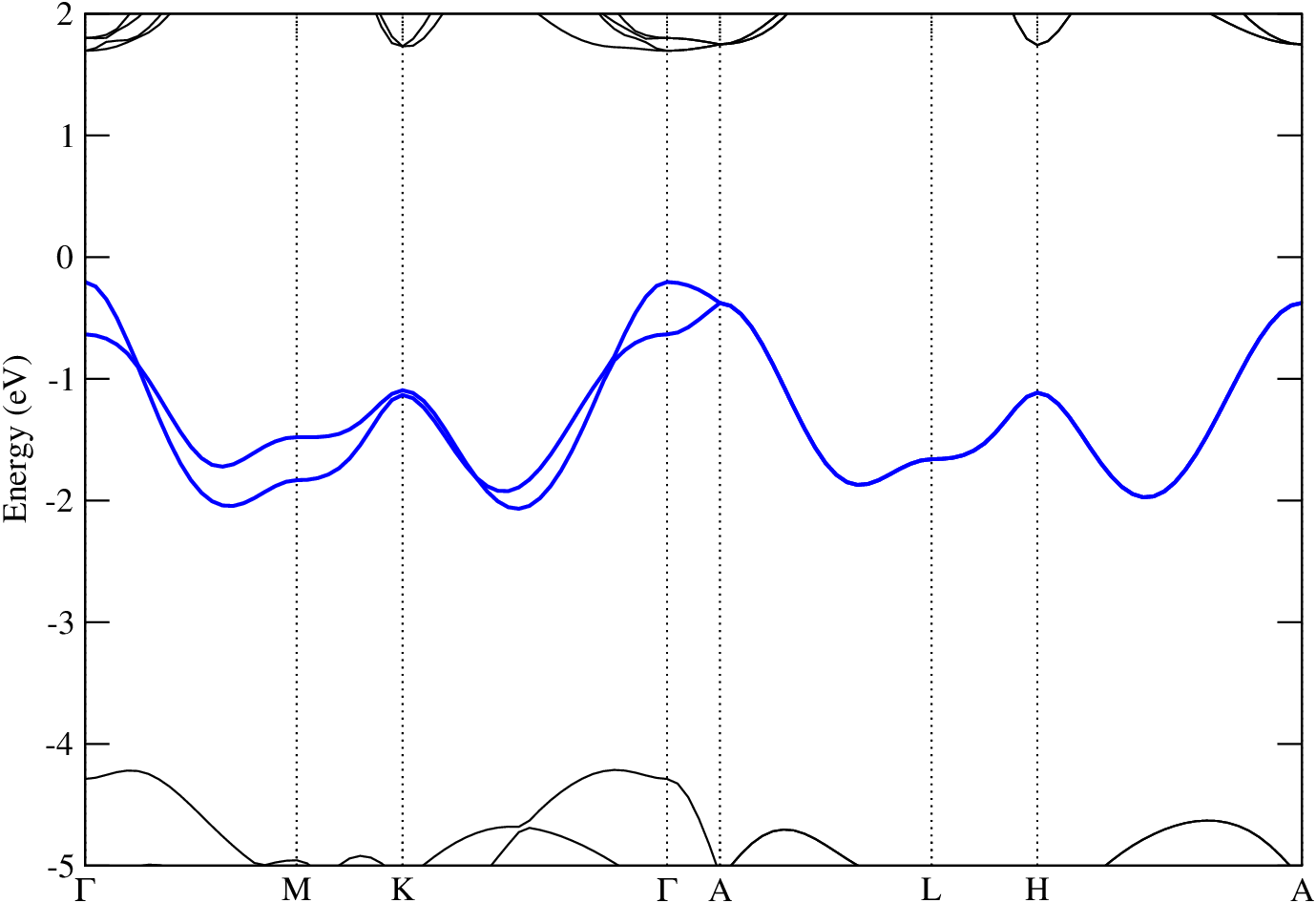}}
\caption{Band structure for $z_O = 0.1263$ (one of the experimental
values, slightly larger than our optimized value of 0.125),
calculated using the code {\it abinit}.\cite{abinit}
Symmetry dictates that the two band stick together along the
A-L-H-A lines ($k_z c = \pi$).  Non-monotonicity of the bands along
the A-L and A-H directions reflects the strong deviation from a
nearest-neighbor hopping behavior, as discussed further in the text.
The crystal field splitting of the Nb $4d$ bands is $\sim$ 4 eV.}
\end{figure}

{\it Electronic structure and tight-binding representation.}  
The band structure of LiNbO$_2$ 
pictured in Fig. 1 is similar to that given earlier by Novikov
{\it et al.}\cite{Novikov94-2} and indicates a Nb $d_{z^2}$ 
bandwidth of 1.9 eV.
The Nb $d_{z^2}$-O $2p$ bands can be fit straightforwardly to a TB model
based on orthonormal
Wannier functions on the two Nb atoms per cell (one Nb per layer).
A full description of the results will be given elsewhere, but we
provide a synopsis here.  There are three important features of the
TB fit that we emphasize here.  First, a good fit requires rather
long range hoppings, up to fourth neighbors within the layer and to
three neighbors in the layers above and below.  Second, with oxygen
ions at their equilibrium position, the second neighbor (in-plane)
hopping amplitude $t_2 \approx$ 100 meV is much larger than the
nearest neighbor hopping $t_1 \approx$ 60 meV.  The smaller value of
$t_1$ may reflect interference between direct Nb-Nb interaction and
the standard O-mediated Nb-O-Nb processes.  The same trend has been
observed for 2H-TaS$_2$,\cite{Wei-arxiv} where the small value of $t_1$ was
traced to phase cancellation in the hopping integral when Wannier
functions are on nearest neighbors.   This $t_2 > t_1$ feature may have
important implications for the microscopic understanding of the
properties of Li$_x$NbO$_2$, since if $t_2$ were the only nonzero hopping,
the system decomposes into three decoupled triangular lattices with
lattice constant $\sqrt{3} a$; $t_1$ then becomes the ``perturbation''
that couples the three sublattices, breaks symmetry and removes
degeneracy.  Thirdly, the nearest
neighbor hopping $t_1$ is very strongly modulated by oxygen displacement.
We find that $t_1$ increases strongly as the O layers ``squash'' against
the Nb layers, as in the $A_g$ Raman mode. 
This modulation may provide the largest contribution to electron-phonon
coupling in this compound.

\begin{table}[bt]
\caption{Born effective charges for LiNbO$_2$, together with
a comparison with NaCoO$_2$ calculated by Li {\em et al.}\cite{Li-Yang}
The angular average $Z^*_{av}$ is also displayed.
Note the unexpected deviations from the formal values Z$^0$
of the effective charges for Li and Nb in the $z$-direction
(larger for Li, smaller for Nb). For O in LiNbO$_2$,
the effective charges are nearly isotropic.  Overall,
the anisotropies are rather similar in NaCoO$_2$,
but somewhat more pronounced.}
\begin{center}
\begin{tabular}{lccccccc}\hline\hline
          &\multicolumn{3}{c}{LiNbO$_2$}&~&\multicolumn{3}{c}{NaCoO$_2$}
\\
                                                   \cline{2-4}\cline{6-8}
          & Li  & Nb  & O &~& Na~ &~ Co & O~    \\\hline
${\mathbf Z}^*_{xx}$ &~~1.10~~&~~2.26~~&~~-1.68~~ &~~&~~0.87~~
        &~~2.49~~&~~-1.68~~   \\
${\mathbf Z}^*_{zz}$&~~1.69~~&~~1.31~~&~~-1.50~~ &~~&~~1.37~~
        &~~0.87~~&~~-1.12~~   \\\hline
${\mathbf Z}^*_{av}$&~~1.30~~&~~1.94~~&~~-1.62~~ &~~&~~1.04~~
        &~~1.95~~&~~-1.49~~   \\\hline
${\mathbf Z}^0$~& +1  & +3  & -2 &~~& +1   & +3 & -2   \\ \hline\hline
 \end{tabular}
\end{center}
\label{tbl:Born}
\end{table}

\begin{figure*}[tbp]
\label{fig:nxcobands}
\resizebox{16cm}{6.5cm}{\includegraphics{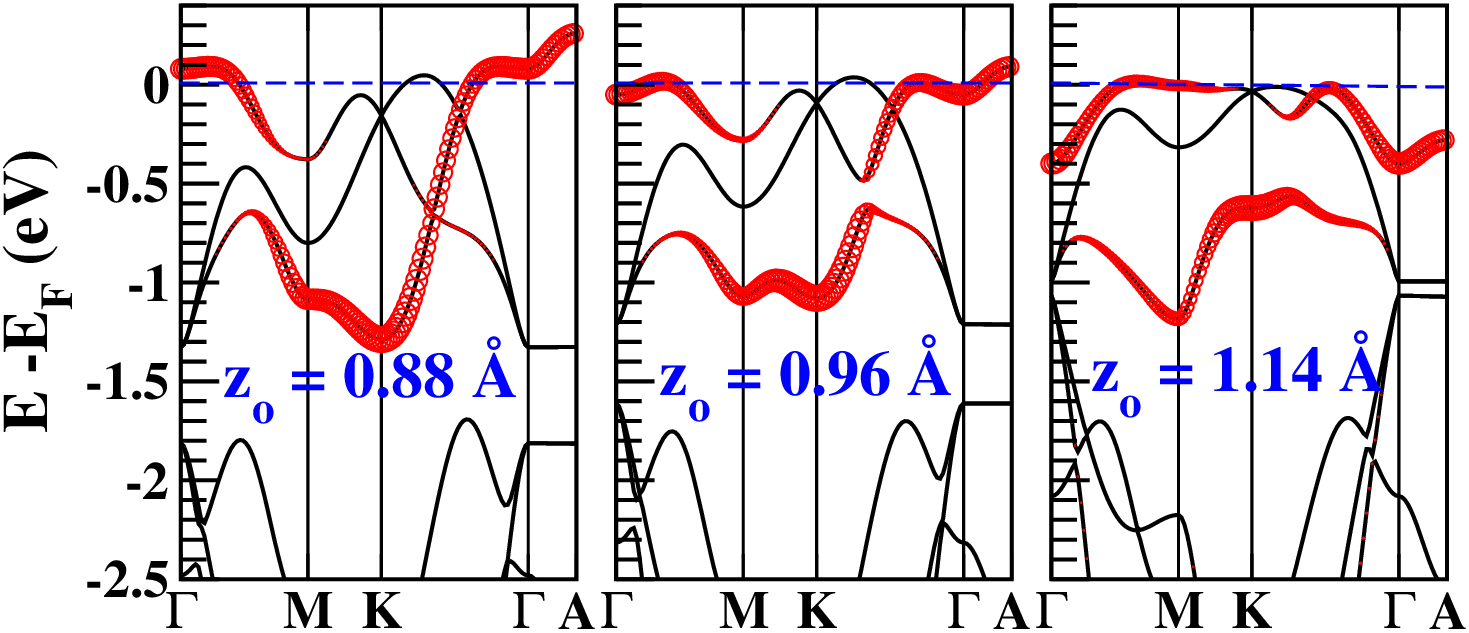}}
\caption{The $t_{2g}$ bands (which lie above -1.5 eV in these figures)
of paramagnetic Na$_{0.5}$CoO$_2$ for
three values of the O height $z_O$.  The value 1.14~\AA~corresponds
to symmetric CoO$_6$ octahedra.  The thickened line emphasizes the
$a_g$ character.  The changes in the $t_{2g}$ bands are discussed in
the text.  Note also the changes in the O $2p$ bands just below
the $t_{2g}$ bands.}
\end{figure*}

{\it Effective charges.}
We have evaluated the Born effective charge tensor as described by 
Gonze and Lee\cite{Gonze-response-2} using the {\it abinit} code.\cite{abinit}
Given in Table \ref{tbl:Born} are the two distinct elements of the effective 
charge tensor for each atom type, calculated in the relaxed atomic structure,  
together with the formal charges.
$Z^*_{xx}$(Li) ($Z^*_{yy}=Z^*_{xx}$) is close to the formal charge of Li
indicating primarily ionic type bonding for motion in the $x-y$ plane, 
consistent with its propensity
for de-intercalation.  The charge tensor for
Li shows similar anisotropy to that of LiBC\cite{Kwan-Woo}, which is
similar structurally and electronically (if some Li is de-intercalated)
to the 40 K superconductor MgB$_2$.  In LiBC $Z_{xx}^*$(Li)=0.81,
$Z_{zz}^*$(Li)=1.46, and
it was concluded that Li
is involved in electronic coupling (not only ionic, but covalent)
between consecutive B-C layers.  Similar Li involvement might be expected
in LiNbO$_2$, and indeed the band structure shows clear effects of
interlayer coupling.
The difference from the formal charges for the Nb ions 
(formally Nb$^{3+}$, O$^{2-}$)
indicate substantial covalent character to the bonding, which appears 
to be especially strong 
for $z$ displacement of the Nb ion.  

The Born effective charges have been reported\cite{Li-Yang} 
for NaCoO$_2$, and since we
investigate O squashing in this compound in the next section, we have included
the NaCoO$_2$ effective charges in Table \ref{tbl:Born} for comparison.
Indeed there are several similarities, as noted in the table caption.

\begin{figure}[tbp]
\label{fig:dos}
\vskip 5mm
\resizebox{8cm}{7cm}{\includegraphics{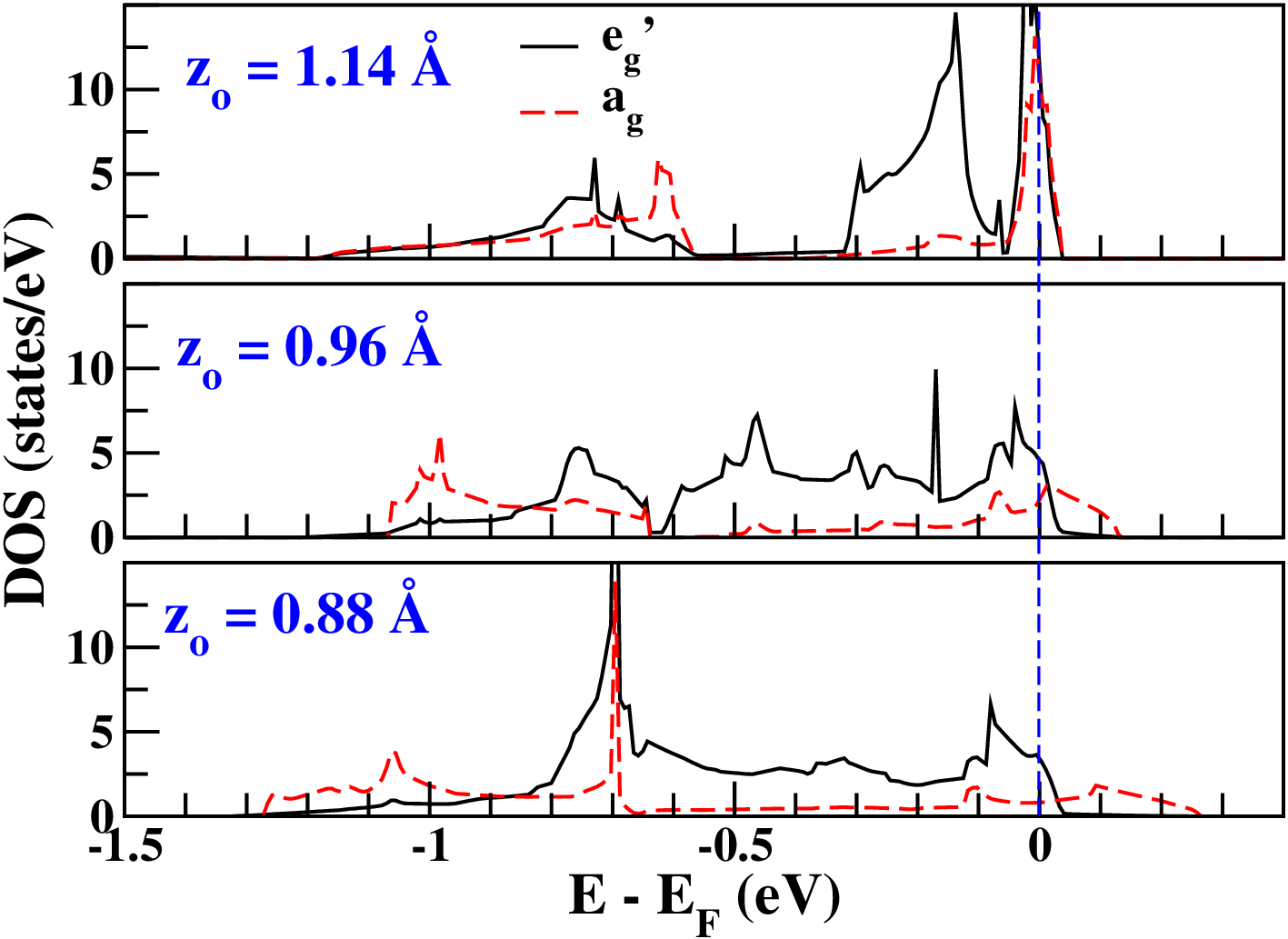}}
\caption{The $a_g$ and $e_g'$ densities of states for the oxygen
heights, and bands, shown in Fig. 2.  Note that the
band widths are identical for the symmetric octahedra
($z_O$=1.14~\AA) but the two-dimensional dispersion already
results in a strongly differing DOS in the upper regions (where,
within a rigid band picture, the doped holes reside).}
\end{figure}

\section{Layered Sodium Cobaltates}
There is already a substantial literature on the electronic structure
of the Na$_x$CoO$_2$ system.  Briefly: the $t_{2g}$ bands are broken
in symmetry by the
layered structure and by the squashing of the CoO$_2$ layers away from
ideal cubic coordination by the six O ions.  The bands are doped
with $1-x$ holes, with all the evidence indicating the holes go, at
least initially, into $a_g$ states rather than $e_g'$ states.  Using
the observed structure, it is found that this results from the somewhat
larger $a_g$ bandwidth, because the band centers remain indistinguishable.

We address here the effect of the height of the O ions above/below the Co layer.
In the calculations, the full-potential nonorthogonal local-orbital 
minimum-basis scheme (FPLO) was used.\cite{fplo}
For specific doping levels and treatments of the Na ions, the height has been
optimized by a number of groups\cite{PZhang,MDJ,JNi,ZLi,KWLee}, revealing
that there is some sensitivity of the O position to the environment.
To clarify the question of the effect of squashing without reference to
a specific doping level, we display in Fig.
2 the $t_{2g}$ bands for O height (from the Co layer)
of 1.14~\AA~
(corresponding to undistorted CoO$_6$ octahedra), 0.96~\AA~(typical value for
intermediate values of $x$), and 0.88~\AA~(the smallest value reported).
For orientation, we note that Johannes {\it et al.},\cite{MDJ}
using the virtual
crystal approximation for the Na concentrations $x$ = 0.3, 0.5, and 0.7,
obtained the heights 0.88, 0.90, 0.93~\AA~respectively.
The corresponding projected densities of states are shown in Fig. 3.
For these calculations we used $x$=0.5, treated within the virtual
crystal model.
To avoid unphysical O-O interactions across the layers as the O layer
position was varied, the $c$ axis was
artificially increased by 20\% for these calculations.

Simple crystal field arguments would suggest: (1) for the cubic octahedron
$z_O$ = 1.14~\AA, the $a_g$ and $e_g'$ DOS should be the same, and (2)
as the O ions are squashed down, the $e_g'$ states should rise relative
to the $a_g$ states.  The first expectation is severely violated in the
region just below $E_F$ due to the dispersion being only two-dimensional
(presuming crystal fields from ions beyond nearest O ions are negligible).
In addition, the effects of squashing are much more complex than suggested
by the crystal field model.  There is minor change in the mean energies of
the $a_g$ and $e_g'$ states (they remain essentially equal, see Fig. 3), the main
change is an {\it increase} in the $a_g$ bandwidth compared to that of
the $e_g'$ states upon squashing. For $z_O$ = 1.14~\AA, doped
holes initially would go equally
into each band.  At the highly squashed end, $\sim$0.4 holes/per Co can go
into the $a_g$ band before encountering the $e_g'$ states.  We emphasize
that this is a model, constrained result; self-consistency and geometrical
relaxation will change the details.  There is also the question of decreasing
interaction with the O $2p$ states upon squashing.  This change, which is
of course also included in the changes shown in Figs. 2 and 3, may affect
the $a_g$ and $e_g'$ states differently.

The changes in the band structure, Fig. 2, are more instructive.
At $\Gamma$, the $a_g$ state is almost 0.5 eV below its maximum for the cubic
octahedron $z_O$=1.14~\AA, the maxima occurring
midway along both $\Gamma$-M and $\Gamma$-K lines.  The additional structure,
and the associated decrease in bandwidth reflects longer range hopping, and
most likely a strong change in the ratio $t_2/t_1$, analogous to the changes
in LiNbO$_2$ but with additional complications due to the presence of the
$e_g'$ bands.  The shift with squashing motion in the $e_g'$ bands is
noticeable not only at $\Gamma$, where the state increases in energy, but also
in the degeneracy at the K point, which rises to the top of the $t_{2g}$
bands for the symmetric CoO$_6$ octahedron.

\section{Summary}
In this paper we have briefly compared and contrasted the electronic structure
of the little-studied layered TMO LiNbO$_2$ to that of Na$_x$CoO$_2$, which has
attracted tremendous since superconductivity was discovered in its hydrate.
Although the active states are quite different, both systems show a strong
change of electronic structure with changes in the TM-oxygen distance.  The
niobate is unusual in having a large second-neighbor hopping amplitude, and
it also presents the attractive simplicity of a single active band on a
triangular lattice.  One of the primary questions to address is whether
electronic correlations are important in the delithiated system, and whether
the origin of superconductivity is of electronic or lattice origin.  

\section{Acknowledgments}
We acknowledge stimulating comments from D. Khomskii and R. J. Cava on the
effect of oxygen ``squashing'' in the Na$_x$CoO$_2$ system, and 
clarification from M. D. Johannes on calculations relating to this question. 
This work was supported by National Science Foundation Grant
DMR-0421810.


\end{document}